\documentclass[12pt,manuscript,authoryear]{aastex}

\usepackage{graphicx}

\begin{document}

\title{The Formation of Fragments at Corotation in Isothermal Protoplanetary Disks}
  
\author{{\it Short Title: Disk Fragmentation ~~~~Article Type: Journal}}

\author{Richard H. Durisen}
\affil{Astronomy Department, Indiana University,
    Bloomington, IN 47405}
\email{durisen@astro.indiana.edu}

\author{Thomas W. Hartquist}
\affil{School of Physics and Astronomy, University of Leeds, Leeds LS2 9JT, UK}
\email{twh@ast.leeds.ac.uk}

\and

\author{Megan K. Pickett}
\affil{Department of Physics, Lawrence University, 
Box 599, Appleton, WI 54912}
\email{megan.pickett@lawrence.edu}

\begin{abstract}
Numerical hydrodynamics simulations have established that disks 
which are evolved under the condition of local isothermality will fragment into 
small dense clumps due to gravitational instabilities when the Toomre 
stability parameter $Q$ is sufficiently low. Because fragmentation 
through disk instability has been suggested as a gas giant planet 
formation mechanism, it is important to understand the physics 
underlying this process as thoroughly as possible. In this paper, we offer 
analytic arguments for why, at low $Q$, fragments are most likely to 
form first at the corotation radii of growing spiral modes, and we 
support these arguments with results from 3D hydrodynamics 
simulations. 
\end{abstract}


\label{firstpage}

\section{Introduction}

The idea that gas giant planets can form rapidly through gravitational
instabilities (GIs) in protoplanetary disks \citep{boss97,boss98},
usually called the ``disk instability'' theory, has now been subjected
to intensive study with numerical hydrodynamics techniques
\citep{pickett98, pickett00, pickett03, nelson98, nelson00, nelson06,
boss00, boss01, boss02, boss03, boss05, boss07, gammie01, mayer02, 
mayer04, mayer07, 
johnson03, rice03, rice05, mejia05, cai06, boley06, boley07}. 
Although all investigators agree that massive, cold protoplanetary disks
fragment into dense clumps under conditions of rapid cooling or local
isothermality, there is not universal agreement that these fragments
are sufficiently long-lived to be considered bound gas giant
protoplanets or that fragmentation will occur under realistic disk
conditions \citep{durisen03, durisen07, boss04, boss05, rafikov05, 
rafikov06, pickett07a}. This important question may
ultimately be decided through simulations alone, but we will gain
greater confidence in the numerical results if we understand the
physics of fragmentation and clump longevity through analytic
arguments, even if only approximate.

As a first step in this direction, we consider the special case of a
disk evolved under a ``locally isothermal" assumption, i.e., the disk
is assumed to maintain its initial temperature at all positions
\citep{boss98,pickett98,nelson98}.  For brevity, most GI researchers
refer to this as an ``isothermal'' disk, even though the temperature
of the disk is not the same at all positions. Local, thin-disk
simulations with a variety of $Q$s by \citet{johnson03} suggest that
fragmentation of isothermal disks occurs for $Q <$ about 1.4, and the
set of all global 3D hydrodynamics simulations for isothermal disks
referenced in the preceding paragraph generally supports the occurrence
of fragmentation for such low $Q$s.
The same set of simulations, taken together, also generally
confirms that disks are subject to growth of nonaxisymmetric GIs for
$Q <$ about 1.7 or so and are stable for higher $Q$s.  
In simulated disks that exhibit
GIs, the instabilities are initiated by the growth from noise of
discrete spiral modes \citep{pickett98,pickett00,pickett03}. Our own
latest isothermal disk simulations, which will be reported in detail
elsewhere \citep{pickett07b}, show three types of behavior: 1) at low
$Q$, {\sl prompt} fragmentation of spiral GI modes as they first
become nonlinear, 2) at intermediate $Q$, {\sl delayed} fragmentation in the
nonlinear regime, probably due to nonlinear interactions of multiple spiral
modes, and 3) at the highest unstable $Q$s, nonlinear spirals which
do not fragment.

In this paper, we explore the idea that prompt fragmentation
in isothermal disks tends to occur near the radius at which a
discrete growing spiral mode corotates with the disk gas (the ``corotation
radius'' or CR). In Section 2, we give simple analytic arguments 
that compression by spiral shocks must not be too large if the 
shock-compressed gas is going to be able to fragment and that the 
smallest compression in a spiral mode should be found near its CR.
In Section 3, we describe 
representative results from a simulation in \citet{pickett07b} where 
prompt fragmentation does in fact occur in the vicinity 
of the CR for a discrete fast-growing mode at low $Q$. The 
simulation allows us to estimate some of the uncertain factors 
in the analysis of Section 2 and verify that the analytic arguments 
do apply. Section 4 summarizes our main conclusions. 

\section{Compression-Induced Stability
Against Fragmentation}

Consider cylindrical coordinates ($r$,$\phi$,$z$) with the rotation axis of 
the disk along the $z$-axis, the disk midplane at $z = 0$, and the sense of 
rotation in the positive $\phi$-direction.
In the absence of spiral structures and shocks, a low-$Q$ disk
has a half-thickness $H \ll r$ in the $z$-direction defined such that 
$\Sigma_1 = 2\rho_1 H$, where $\Sigma_1$ is the surface density 
and $\rho_1$ is the midplane density. 
We shall assume that the initially fastest-growing discrete spiral mode 
develops more rapidly than any other instability which may give rise to 
fragments. A spiral shock at $r$ associated with the mode then sweeps 
material into a sheet of vertical height $H$ with thickness $L$ normal 
to the spiral shock front. Because the gas responds isothermally to the
shock, there is no vertical jump behind the shock, and the vertical height
of the disk remains essentially the same on both sides of the shock 
\citep{bd06}.

We will analyze the gravitational stability of the swept-up gas by 
assuming that it is well approximated by a thin, 
plane-parallel sheet. This requires $L \ll r$ and
\begin{equation}
L < 2\alpha_1 H,
\end{equation}
where $\alpha_1$ is a constant of order unity which accounts for 
uncertainty about how much smaller $L$ must be than 2$H$ for our
subsequent analysis to be at least approximately valid. Throughout, 
we will introduce similar factors, each of which is presumably of order 
unity and designated by an $\alpha_{j}$, where $j$ is an index. 
They account for uncertainties associated with our 
assumptions.

If the density $\rho_2$ in the swept-up sheet can be considered roughly 
constant over scales of order $L$, which seems reasonable, then 
potential energy per unit area of the gas sheet is 
\begin{equation}
E_G = -\pi G{\rho_2}^2 L^3/6,
\end{equation}
where $G$ is the gravitational constant. There are several effects that can
counter the tendency of self-gravity to cause fragmentation,
including thermal pressure, tidal gravitational stresses, and shearing motions. 
The Virial Theorem suggests that an infinite, isothermal, plane-parallel,
ideal gas slab will be stable against fragmentation due to its thermal 
energy alone if $|E_G| < 2E_T$, where $E_T$ is the total thermal kinetic 
energy of the slab per unit area. This gives
\begin{equation}
\pi G{\rho_2}L^2 < 18\alpha_2{c_{\rm i}}^2,
\end{equation}
where $c_{\rm i}$ is the isothermal sound speed and $\alpha_2$ accounts for 
the uncertainty due to deviations from a thin, uniform slab geometry. Stability 
condition (3) simplifies to 
\begin{equation}
{\Sigma_{\rm slab}}^2/\rho_2 < 18\alpha_2{c_{\rm i}}^2/\pi G,
\end{equation}
where $\Sigma_{\rm slab} = \rho_2 L$ is the surface mass density of
the swept up slab of gas in one of the spiral arms. If the spiral mode has $m$
equally-spaced arms that together have swept up a fraction $f$ of the 
material in each annulus of the disk at the time of fragmentation, then
\begin{equation}
\Sigma_{\rm slab} = 2\pi rf\rho_1/m.
\end{equation}

From equation (4), we see that, for a fixed value of $\Sigma_{\rm slab}$,
fragmentation is inhibited by large values of $\rho_2$. In other words,
strong compression by spiral shocks $suppresses$ fragmentation. 
This somewhat counterintuitive result is peculiar to isothermal slabs. 
Compression by the spiral shock is least near the radius at which the
spiral pattern corotates with the disk orbital motion. At significant 
distances from CR, shock compression will be strong and will suppress
fragmentation. A major aim of this section is to estimate the minimum 
distance away from the CR at which this suppression is effective. 

We first examine criterion (1) by noting that
\begin{equation}
L = 2\pi rf\rho_1/m\rho_2.
\end{equation}
If the shock is strong and isothermal, then
\begin{equation}
\rho_2/\rho_1 = {v_{\rm sh}}^2({\rm sin}\psi)^2/{c_{\rm i}}^2,
\end{equation}
where $v_{\rm sh}$ is the $\phi$-direction speed of the spiral shock front 
relative to the pre-shock material and $\psi$ is the angle to the normal 
of the spiral shock front made by the fluid streamlines coming into the shock. 
At a distance $\delta r$ from the CR of the spiral mode,
\begin{equation}
v_{\rm sh} = -3\alpha_3\Omega\delta r/2,
\end{equation} 
where $\delta r/r \ll 1$ and $\Omega$ is the disk's undisturbed angular 
frequency of rotation. The factor $\alpha_3 = 1$ for a disk in Keplerian 
rotation but may differ from unity for disks with significant 
self-gravity or pressure support. 
We define $\alpha_4$ such that
\begin{equation}
H = \alpha_4 c_{\rm i}/\Omega.
\end{equation}
Using (6) through (9), we find that the thin-sheet condition (1) is satisfied if
\begin{equation}
\left({\delta r}\over{H}\right)^2 
> \left.{4\pi r}f\over{9\alpha_1{\alpha_3}^2{\alpha_4}^2({{\rm sin}\psi})^2 mH}\right..
\end{equation}

Now let us consider criterion (3) for compression-induced suppression of 
fragmentation. For disk gas that is strictly isothermal, the \citet{toomre64} 
stability parameter $Q_{\rm i}$ is 
\begin{equation}
Q_{\rm i} = c_{\rm i}\kappa/\pi G\Sigma_1,
\end{equation}
where the epicyclic frequency $\kappa$ is $[d(r^2\Omega)^2/r^3dr]^{1/2}$
\citep{binney85}.
We set 
\begin{equation}
\kappa = \alpha_5\Omega,
\end{equation}
so that $\alpha_5 = 1$ for a Keplerian disk.
Using (4), (5), (7), (8), (9), (11), and (12) plus the definition of $\Sigma_1$,
we get that condition (3) for suppression of fragmentation is satisfied if
\begin{equation}
\left({\delta r}\over{r}\right)^2 
> \left.{4\pi^2\alpha_5}f^2\over{81\alpha_2{\alpha_3}^2{\alpha_4}
({{\rm sin}\psi})^2 m^2Q_{\rm i}}\right..
\end{equation}
Together inequalities (11) and (13) show that, for sufficiently large values  
$\delta r$, both conditions (1) and (3) are met. In other words, away from the
CR of a growing spiral mode with $m$ arms, shock compression due to the 
mode creates a thin sheet of gas which is stable against fragmentation. 
If prompt fragmentation is to occur, it must do so in the vicinity of the CR. 
According to (11) and (13), as $m$ decreases, the region 
around the CR where the post-shock slab is not stabilized increases. Whether 
prompt fragmentation will occur thus depends on what mode with what 
corotation radius is likely to reach nonlinear amplitude first.

So far, there is no rigorous analytic theory which predicts the number of 
arms for the fastest growing nonaxisymmetric mode of a gravitationally 
unstable disk. However, a WKB analysis for axisymmetric modes of  thin disks 
with $Q < 1$ \citep{toomre64,binney85} yields a wavelength for the fastest 
growing ring-like mode of
\begin{equation}
\lambda_{\rm f} \approx 2.2\pi^2G\Sigma_1/\kappa^2.
\end{equation}
Comparisons between simulations and equation (14) \citep{durisen03} suggest
that, for disks unstable to nonaxisymmetric modes, the fastest growing
nonaxisymmetric mode has a number of arms given by
\begin{equation}
m_{\rm f} = \alpha_6\pi r/\lambda_{\rm f} 
= \alpha_4\alpha_5\alpha_6Q_{\rm i}r/2.2H,
\end{equation}
where $\alpha_6 \sim 1$. 
Disks do not break up directly into fragments of this size.
Instead, the simulations show that the mode grows to nonlinear 
amplitude as a spiral wave. In the case of prompt fragmentation, 
fragments appear first as a smaller-scale instability in the dense 
post-shock region associated with the nonlinear wave.
If we set $m = m_{\rm f}$ in (10) and (13), the 
thin-sheet criterion (1) becomes 
\begin{equation}
\left({\delta r}\over{H}\right)^2 
> \left.{8.8\pi f}\over{9\alpha_1{\alpha_3}^2{\alpha_4}^3\alpha_5\alpha_6
({{\rm sin}\psi})^2 Q_{\rm i}}\right.
= \left.{3.1\zeta}\over{\alpha_1Q_{\rm i}}\right.,
\end{equation}
and criterion (3) for compression-induced stability against fragmentation
becomes
\begin{equation}
\left({\delta r}\over{H}\right)^2 
> \left.{0.24\pi^2f^2}\over{\alpha_2{\alpha_3}^2{\alpha_4}^3
\alpha_5{\alpha_6}^2({\rm sin}\psi)^2 Q_{\rm i}^3}\right. 
= \left.{2.4f\zeta}\over{\alpha_2\alpha_6Q_{\rm i}^3}\right.,
\end{equation}
where 
\begin{equation}
\zeta = f/{\alpha_3}^2{\alpha_4}^3\alpha_5\alpha_6({{\rm sin}\psi})^2.
\end{equation}

The similar numerical quantities on the right hand sides of conditions (16) and (17) 
suggest that, whenever the spiral shock compresses gas into a thin slab, it will be stable
against fragmentation. However, this also means that the thinness and strong shock 
assumptions required for use of relations (3) and (7) break down together simultaneously 
near the CR. We cannot invert the inequalities to obtain conditions for instability because
there is at least one other stabilizing effect, namely, shear.
So, strictly speaking, our analysis does not tell us whether or when 
prompt fragmentation $does$ occur, but it implies that, $if$ prompt fragmentation 
occurs, then it must happen in the vicinity of the CR for conditions under which 
the inequalities are reversed. We can see from the $Q_{\rm i}$-dependence of (17) that this
is more likely to happen for low $Q_{\rm i}$. 

\section{Comparison with Simulations}

\subsection{Methods and Initial Conditions}

Adopting the same basic star/disk model and 3D hydrodynamics code 
used by \cite{pickett98}, we have computed a series of 
locally isothermal simulations of disks with different, but constant, values of $Q$. 
The $r,z$-resolution is the same as in \cite{pickett98}, but the 
$\phi$-direction resolution is increased from 128 to 512 azimuthal cells in 2$\pi$ radians.
This seems sufficient to resolve fragmentation in modes of moderate $m$ 
and to satisfy the \cite{truelove97} and \cite{nelson06} criteria for avoiding purely 
numerical fragmentation prior to the onset of fragmentation. 
So far, the series of simulations includes $Q =$ 1, 1.15, 1.25, 1.35, 1.5, 1.6, and 1.7. 
For reasons explained in Section 3.2 below, $Q$ here, without the subscript ``i", 
is computed using (11) but replacing the isothermal sound speed by the adiabatic sound 
speed for an ideal gas with ratio of specific heats $\gamma = 5/3$. Consequently, 
\begin{equation}
Q = (5/3)^{1/2}Q_{\rm i} = 1.29Q_{\rm i}
\end{equation}
The star/disk model is the same massive ``stubby'' disk used in \cite{pickett98,pickett00}, 
with disk-to-star mass and radius ratios of about $M_d/M_s =$ 0.25 and $R_d/R_s =$ 7.1, 
respectively. In this case, however, in order to avoid the severe cost of 
the Courant time-step limitation near the rotation axis, the disk is detached 
from the star using the localized cooling method 
described in \cite{pickett03}, and the stellar mass distribution and gravitational 
potential are frozen. 

\subsection{Prompt Fragmentation at Corotation}

In the full $Q$-survey, which will be reported in greater detail elsewhere 
\citep{pickett07b}, the initial disk is given a small amplitude, random, cell-to-cell 
density perturbation. This allows fast growing modes to organize themselves and
grow in a few dynamic times. The three behaviors
described in the Introduction are found over the following ranges: 1) $1.0 \le Q \le 1.25$:
Prompt Fragmentation. Fragments appear as soon as growing modes reach 
nonlinear amplitude. 2) $1.35 \le Q \le 1.6$: Delayed Fragmentation. 
Discrete modes do not fragment as soon as they reach nonlinear amplitude; instead, the 
appearance of dense fragments is delayed by several to many pattern rotations and
appears to be associated with nonlinear interactions of different patterns or arms. 
3) $Q = 1.7$: Nonfragmenting Spiral Arms.  The disk is unstable and develops strong 
spiral arm structure, but no fragmentation occurs over the duration of a long simulation. 
We have not yet tested the upper bound of GI stability for this star/disk model, but we 
do know from \cite{pickett98} that a $Q = 2.0$ disk is stable against growth of 
nonaxisymmetric structure, so the $Q$-limit for onset of GIs is somewhere between 
1.7 and 2.0. Clearly, the $Q$ below which fragmentation occurs for these isothermally 
evolved disks is between 1.6 and 1.7. We compare this fragmentation limit with 
that found by \cite{johnson03} in Section 3.2 below. 

For the low $Q$s of interest here, where prompt fragmentation occurs, many modes grow 
rapidly at once, and it is difficult to discern which mode or pattern period is associated 
with a given fragment or set of fragments. So, for this paper, we apply Fourier analysis
techniques \citep{pickett98,pickett00} to the $Q = 1.25$ case and determine, as best
as we can, that the fastest-growing mode is a particular $m = 4$ mode. We then run an additional 
simulation with a pure, low-amplitude $\delta\rho/\rho \sim {\rm cos}4\phi$ density 
perturbation over the radial range where the fastest growing $m =4$ pattern was 
detected in the random perturbation simulation. As a cautionary note, we point out that 
it is difficult to be sure we have truly isolated the {\it most} unstable mode, 
because many modes grow with similarly fast growth rates. A midplane greyscale of 
the disk close to the moment of prompt fragmentation in the run with the pure cos4$\phi$ 
hit is shown in Fig.~1, where it is apparent that the fragments do indeed appear near 
the CR of the growing mode.

We can use the simulation to estimate some, but not all of the various 
parameters that enter into our analysis in Section 2. The rotational shear of this 
massive disk is somewhat non-Keplerian, so that $\alpha_3 \approx$ 0.9 and 
$\alpha_5 \approx$ 0.9. Using $\Sigma_1$ and the midplane values of 
$\rho_1$, $c_{\rm i}$, and $\Omega$ near the CR in (9), we find that 
$\alpha_4 \approx$ 0.6. Use of these $\alpha_{j}$s and the known values of $Q_{\rm i}$, 
$H$, the $r$ of CR, and $m_f = 4$ in (15) gives $\alpha_6 \approx$ 0.7. 
It is difficult to measure $\psi$ in our Eulerian code, but it is relatively
easy to determine the pitch angle of the spirals to be about 20 degrees prior to 
the onset of fragmentation. So we use $\psi \approx 20$ degrees as a crude estimate, 
although the pre-shock $\psi$ is likely to be somewhat larger than the pitch angle 
due to radial motion of the fluid as the wave becomes nonlinear. The fraction of mass $f$ in the 
spirals just prior to fragmentation is also difficult to determine precisely but appears 
to be about 0.2. Using these values in (18) gives $\zeta \approx$ 16. If we assume 
$\alpha_2 = 1$, then (17) becomes
\begin{equation}
\left({\delta r}\over{H}\right)^2 
> \left.{11\over{Q_{\rm i}^3}}\right.,
\end{equation}
Unfortunately, it is difficult to determine the true $\delta r$ for the parts of the spirals 
that go into the dense fragments, and so it is difficult to verify that (20) is 
precisely satisfied. Nevertheless, for $Q$ = 1.25 ($Q_{\rm i} = 0.97$), (20) gives 
$\delta r \sim$ 3 to 4$H$ for the edge of the stabilized region, 
or about six cell widths from the CR.  
The clumps are well within this distance from the CR
in disk midplane images near the time of fragmentation.

Other aspects of our analysis can also be checked against the
simulation results.  For instance, the thinness of the arms in the
midplane view of Fig.~1 suggests that condition (1) is satisfied, 
but this observation does not give us much insight into the proper value of
$\alpha_1$.  Finally, consider $m_f$. Both $Q_{\rm i}$ and $H$ in (15) vary
linearly with $c_{\rm i}$, while other parameters involved in $Q_{\rm i}$ and $H$,
like $\kappa$, $\Omega$, and $\Sigma_1$, are determined primarily by
the disk's radial structure, which does not vary much with $c_{\rm i}$ for
cool disks.  So we expect $m_f$ to be relatively insensitive to $Q$
until the disk becomes hot enough for radial pressure gradients to
play some dynamic role. In fact, our set of simulations with varied
$Q$ indicate that $m = 4$ is the fastest growing mode for all $Q$ between 1.15
and 1.5 \citep{pickett07b}. The $Q = 1.0$ case has so many fast growing modes 
with various $m$s, including $m = 4$, that we cannot determine 
which of them grows most rapidly.

\subsection{Fragmentation Criteria}

Comparison of fragmentation criteria derived by different authors involves 
some discussion about how $Q$ is evaluated. 
For our disk models, we use (11) but with the adiabatic sound speed 
 \citep{pickett98, pickett00, pickett03}.  
One of the goals of our body of work has been to 
consider the same basic equilibrium disk models evolved under different
assumptions about the equation of state of perturbed fluid elements. 
Because the models are derived from isentropic equilibrium configurations,
it makes sense to use the adiabatic sound speed to compute a common 
reference $Q$ that uniquely designates individual disk models within
a set of related models. For the purposes of comparison with other work, 
however, we need to compute $Q$ the same way other authors do, to the 
extent that we can. \citet{johnson03} and \citet{mayer04} use what we 
call here $Q_{\rm i}$. \citet{johnson03} report fragmentation if and only
if $Q_{\rm i} <$ about 1.4; the \citet{mayer04} paper suggests a somewhat 
larger value. We find fragmentation setting in for $Q <$ 1.7 or,
equivalently, for $Q_{\rm i} <$ about 1.32, which is in reasonable 
agreement with these other works, especially considering that the precise
limit in global simulations probably depends somewhat on the detailed 
structure of the disk. 

Dynamically, it may seem obvious that the isothermal sound speed is
the appropriate one to use in this case. However, the perturbations in 
our simulations are {\it locally} isothermal, i.e., the temperatures are 
fixed spatially throughout the grid. Sound waves traveling in the $\phi$ 
direction are truly isothermal, but waves moving in the $r$ or $z$ directions 
are not. Thus, it is not absolutely clear what sound speed should be used 
in our approximate stability relations, nor how truly ``isothermal" the spiral 
shocks will be. This uncertainty is not explicitly accounted for by any of the
$\alpha_j$s in our analysis. It also adds an unknown level of uncertainty
when comparing gird-based locally isothermal simulations with SPH
simulations in which the temperatures of the particles are kept
fixed instead and thus may more accurately represent 
fluid elements that are evolving isothermally. 

\section{Conclusions}

We have used simple anayltic arguments to estimate where fragments 
can first appear in a low-$Q$ gravitationally unstable disk when the 
spiral shocks are locally isothermal.  Assuming that 
the disk behavior is dominated by a single, coherent spiral mode and 
that the dense gas behind the spiral shock 
can be well approximated as a thin slab behind the shock, 
we have shown that compression from isothermal shocks 
will stabilize the slab against fragmentation 
away from the CR. The minimum distance from the CR at which 
the stabilization becomes effective gets larger for smaller $Q_{\rm i}$. 
So the analysis suggests
that, if fragments are going to form promptly upon nonlinear growth of 
a mode, then this will happen first near the CR of the mode and only 
for low enough $Q_{\rm i}$. The result is prompt 
formation of high density blobs near corotation in each spiral arm 
of the dominant pattern as it becomes nonlinear.
A locally isothermal 3D hydrodynamic simulation of a well-studied 
star/disk model with $Q = 1.25$ ($Q_{\rm i} = 0.97$) 
confirms this prediction. We plan to attempt similar analyses in the 
future to understand fragmentation under different conditions,
including delayed fragmentation and fragmentation 
in radiatively cooled disks. We suspect that, as in the isothermal
case, prompt fragmentation in radiatively cooled disks 
will tend to occur under more extreme
conditions than fragmentation itself. In other words, if fragmentation
generally occurs when $t_{cool}\Omega < c_{\gamma f}$ 
\citep{gammie01,rice03, rice05, mejia05}, where $c_{\gamma f}$ 
is a constant that depends on the adiabatic index $\gamma$ and 
where $t_{cool}$ is the time it takes the disk to radiate away its 
internal energy, then we would expect prompt fragmentation to 
occur when $t_{cool}\Omega < c_{\gamma p}$, where 
$c_{\gamma p} < c_{\gamma f}$.

In closing, we would like to reiterate a crucial point we have made 
elsewhere \citep{pickett98,pickett00,pickett03,durisen03,mejia05,
pickett07a}. The occurrence of fragmentation in disks does not 
in itself demonstrate that protoplanet formation by disk instability is
really possible. The key issues are whether conditions for disk 
fragmentation are achieved in real disks \citep{rafikov05, 
rafikov06, cai06, boley06, boley07} and whether dense 
clumps, if they do form, are able to evolve into protoplanets
over many orbital periods \citep{pickett07a}.  Although our criteria 
suggest where potential precursors to gas giant planets {\it may} appear 
under certain restrictive conditions, they do not say anything about the 
longevity of the clumps once formed.

\section*{Acknowledgments}

We would like to thank S. Slavin for assistance with the computations,
an anonymous referee for helpful comments,
and C.F. Gammie for useful correspondence. This research was 
supported in part by NASA grants NAG5-11964, NNG 05-GN11G, 
and NAG5-10262 and by PPARC Visitors grants.

\newpage

\begin{center}
{\large\bf Figure Captions}
\end{center}

{\bf Figure 1.} Midplane mass density grayscale.  Shown is a grayscale intrepretation of the midplane
mass density at the time when fragments first appear in the 3D hydrodynamics simulation of a 
$Q = 1.25$ disk given a cos4$\phi$ initial perturbation.  The greyscale spans six orders of 
magnitude in density.  The circle indicates the location of the corotation radius (CR).  Note that 
the first clumps to appear are quite close to the CR for the four-armed mode.

\begin{figure}
\includegraphics[width=17cm]{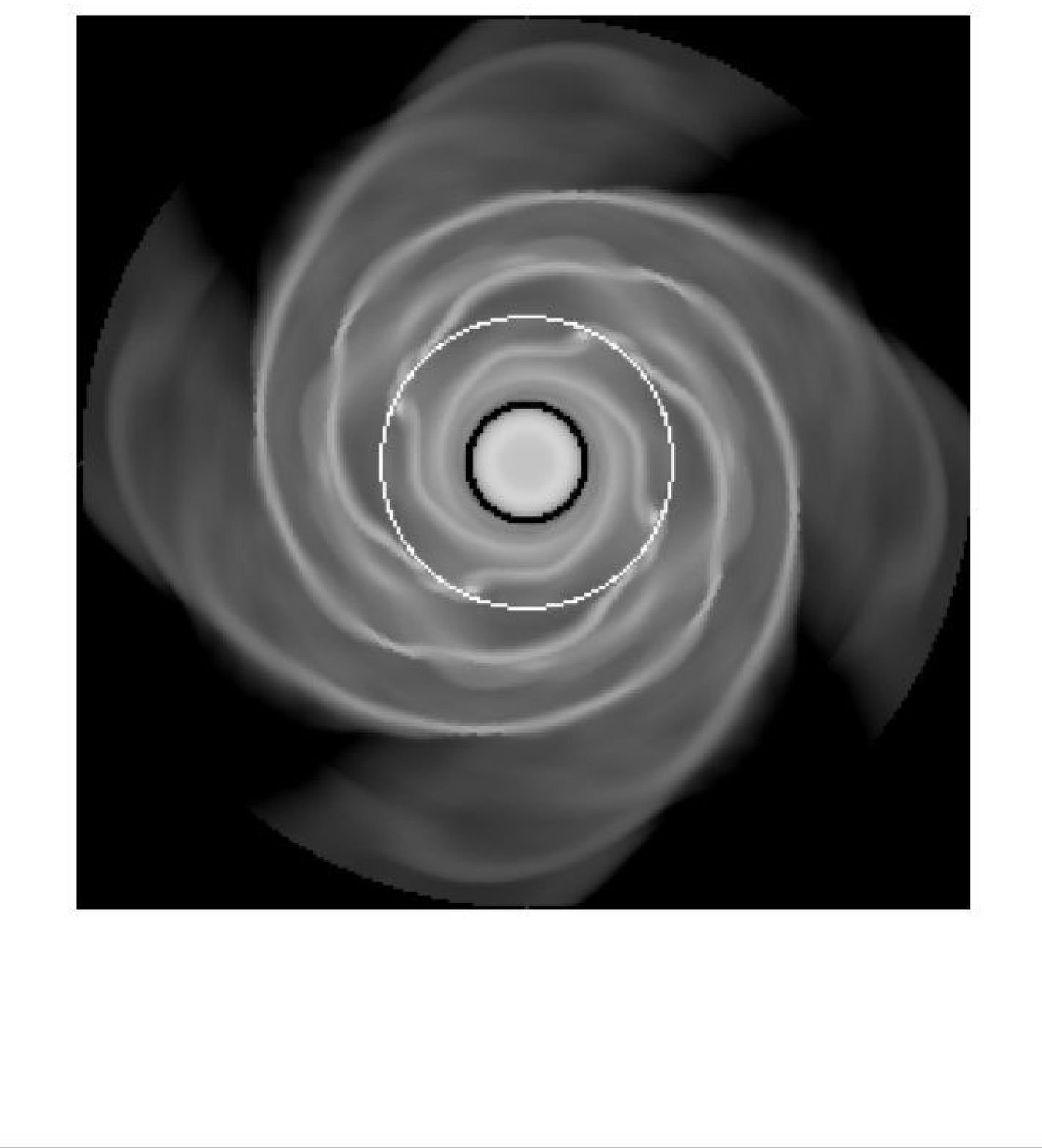}
\end{figure}


\label{lastpage}

\end{document}